\documentclass[a4paper]{article}
\usepackage{17lomcon} 
\usepackage{cite} 
\usepackage{epsfig} 
\usepackage{epstopdf}
\usepackage{color}
\usepackage{graphicx}
\graphicspath{{./figures/},{./}}
\ifx\pdfoutput\undefined
\usepackage[dvips,bookmarks]{hyperref} 
\else
\usepackage{hyperref} 
\fi
\newcommand{\beq}{\begin{equation}}
\newcommand{\eeq}{\end{equation}}
\newcommand{\bea}{\begin{eqnarray}}
\newcommand{\ena}{\end{eqnarray}}
\newcommand{\etal}{{\em et al.}}
\def\PAMELA{{\sc Pamela}}
\def\AMS{{\sc Ams-02}}
\newcommand{\aap}{Astron. {\&} Astrophys.}
\newcommand{\apj}{Astrophys. J.}
\renewcommand{\thepage}{\textcolor{black}{{Proceedings of the 17th Lomonosov Conference} -- {August 22, 2015} -- \arabic{page}}}
\begin{document}
\title{INDIRECT DARK MATTER SEARCHES IN THE LIGHT OF THE RECENT AMS-02 OBSERVATIONS
\vskip 0.25cm
LAPTH-Conf-020/16}
\author{Pierre Salati \email{pierre.salati@lapth.cnrs.fr}}
\affiliation{LAPTh, Universit\'e de Savoie Mont Blanc, CNRS;\\
9 Chemin de Bellevue, B.P.110, Annecy-le-Vieux F-74941, France}
%
\date{}
\maketitle

\begin{abstract}
If the astronomical dark matter is made of weakly interacting, massive and stable species, it should annihilate
on itself into particles. This process should produce rare antimatter cosmic rays and lead to distortions in their
energy distributions. The AMS-02 spectrometer has been measuring them with unprecedented accuracy. It is timely
to investigate if anomalies have been found in the positron and antiproton spectra and if so, if they indirectly point
toward the presence of DM particles annihilating inside the Milky Way.
\end{abstract}

\section{Cosmic rays as an indirect probe for dark matter}

The universe contains a substantial fraction of its mass under the form of the so-called astronomical dark matter (DM),
a pressureless component found inside galaxies \cite{1980ApJ...238..471R,1979A&A....79..281B}, clusters of
galaxies \cite{1933AcHPh...6..110Z} and on cosmological scales. The recent observations of the Planck
satellite~\cite{Ade:2013zuv} have confirmed the picture of a flat universe filled with dark energy (68.3\%),
dark matter (26.8\%) and baryons (4.9\%). According to this standard lore, the astronomical dark matter cannot
be made of baryons and its nature is still unknown.
Many solutions have been proposed for the last three decades. Among the numerous possibilities, a particular candidate
under the form of a weakly interacting massive particle dubbed WIMP has attracted much attention. This species is
naturally present in most extensions of the standard model of particle physics. It is stable by conservation of a quantum
number, such as R-parity in supersymmetry or the momentum along the extra-dimensions in Kaluza-Klein inspired models.
It interacts with its surroundings and annihilates on itself through typically weak interactions. The crucial consequence,
that makes WIMPs so interesting, is that they are produced during the Big Bang with a relic abundance close to the Planck
value of $\Omega_{\rm DM} h^{2} = 0.1196 \, \pm \, 0.0031$. For this to happen, the annihilation cross
section $\langle \sigma v \rangle$ should be close to the canonical value of $3 \times 10^{-26}$ cm$^{3}$ s$^{-1}$.

The searches for WIMP-like DM have developed along three directions.
First, WIMPs could be produced at colliders such as the LHC and appear in missing energy events. An abnormally large
rate of gluon monojets or single gauge boson events could be explained by the fusion of quark-antiquark pairs
into pairs of DM particles.
A second line of research, called direct detection, is based on the potential collisions of WIMPs on a terrestrial instrument.
As an impinging DM species collides on a nucleus, a kinetic energy of a few tens of keV is transferred to the later. Current
techniques are sensitive to tiny recoil energies and the background is now strongely suppressed, making that technique
particularly promising.
%
Finally, WIMPs could also be indirectly detected through the particles which they would produce by annihilating inside the
Milky Way. Although DM annihilation is a marginal process today, it is a potential source of high-energy photons,
neutrinos and rare antimatter particles such as positrons $e^{+}$, antiprotons $\bar{p}$ or even antideuterons $ \bar{D}$
through the set of reactions
\beq
\chi + \chi \; \to \; q + \bar{q} \, , \, W^{+} + W^{-} \, , \, \ldots \; \to \;
\bar{p} , \bar{D} , e^{+} , \gamma \; \& \; \nu \;\; .
\label{indirect_reac}
\eeq
Antimatter cosmic rays are already manufactured by conventional astrophysical processes. The dominant mechanism is the
spallation of primary high-energy protons and helium nuclei on the gas of the Galactic plane. Positrons could also be accelerated
by highly-magnetized neutron stars called pulsars.
The messengers of DM annihilation are expected to generate distortions in the signals detected at the Earth or to appear in the
$\gamma$-ray sky as hot spots with no optical counterpart -- see the review~\cite{Lavalle:2012ef} for more details.

In this presentation, I will be concerned with DM indirect searches in the light of the latest AMS-02 observations. The AMS-02
spectrometer has been measuring the fluxes of charged cosmic rays with unprecedented accuracy. Should there be WIMPs
in the Galaxy, their annihilation products could leave imprints in the energy spectra of antimatter cosmic rays.
Detecting an excess requires to know the background though. Modeling the transport of charged cosmic rays inside the Milky
Way is of paramount importance insofar as the searched signals come out actually as deviations from the conventional
astrophysical spectra. These need to be determined as precisely as possible.
The propagation of charged particles inside the Galactic magnetic field is understood as a mere diffusion process where the
cosmic ray species collide on the turbulent knots of the field. The magnetic halo can be seen as a slab a few kiloparsecs thick
in the middle of which the Galactic disk lies, with its stars and gas. The particles can also be driven apart from the disk through
Galactic convection. Above a few GeV, diffusion and energy losses are the dominant mechanisms.
Primary cosmic rays are interstellar nuclei and electrons of the disk that are accelerated to very high energies by supernova driven
shock waves. As they diffuse throughout the magnetic halo, they can interact on the disk to produce secondary species, such as
positrons and antiprotons, generating a background to the exotic DM signals.

\newpage
\section{The cosmic ray positron excess -- Dark matter versus pulsars}
\label{sec:positrons}
%
\begin{figure}[t!]
\begin{minipage}{6cm}
\centering
\includegraphics[width=6cm]{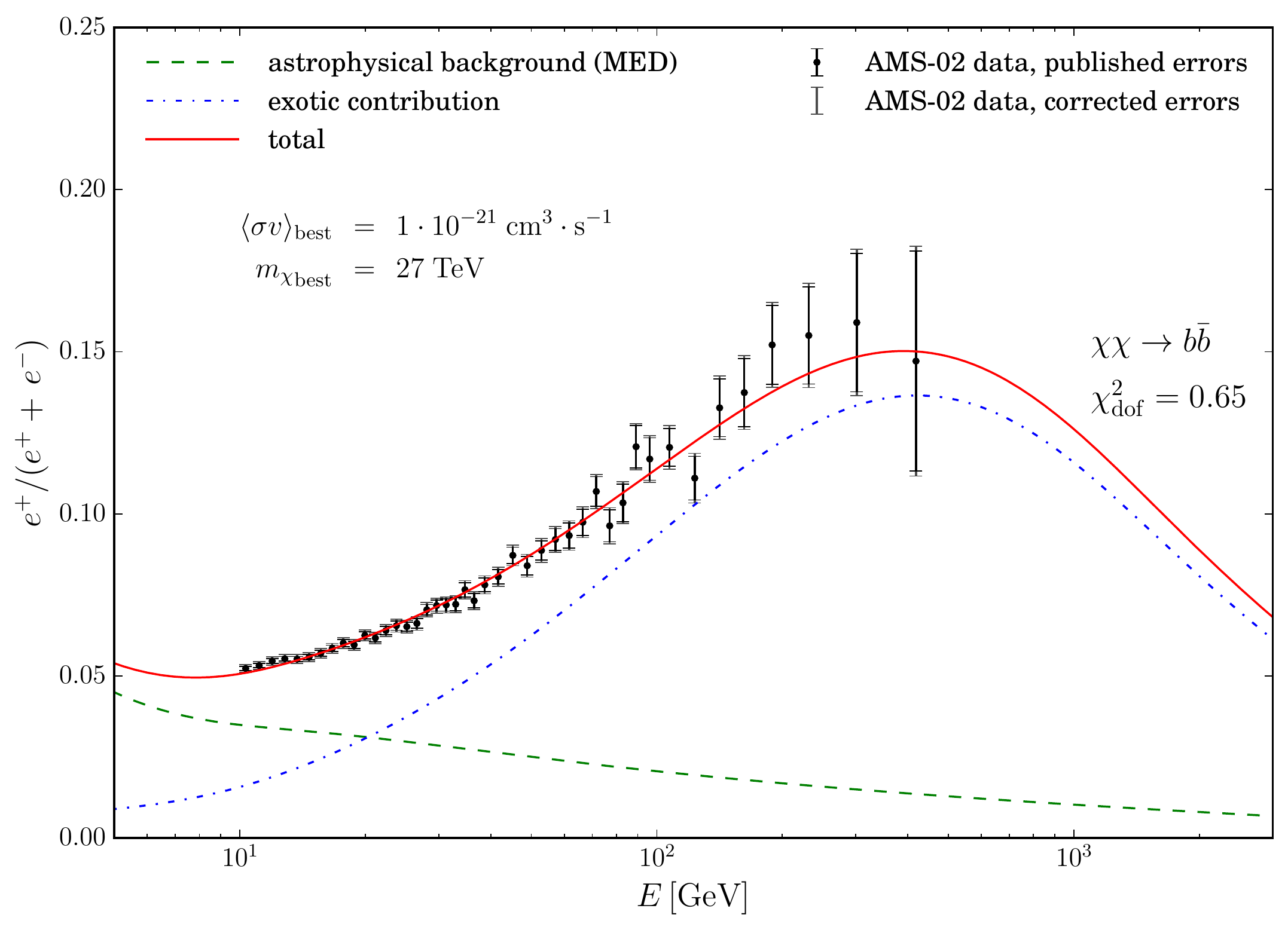}
\end{minipage}
\hfill
\begin{minipage}{6cm}
\centering
\includegraphics[width=6cm]{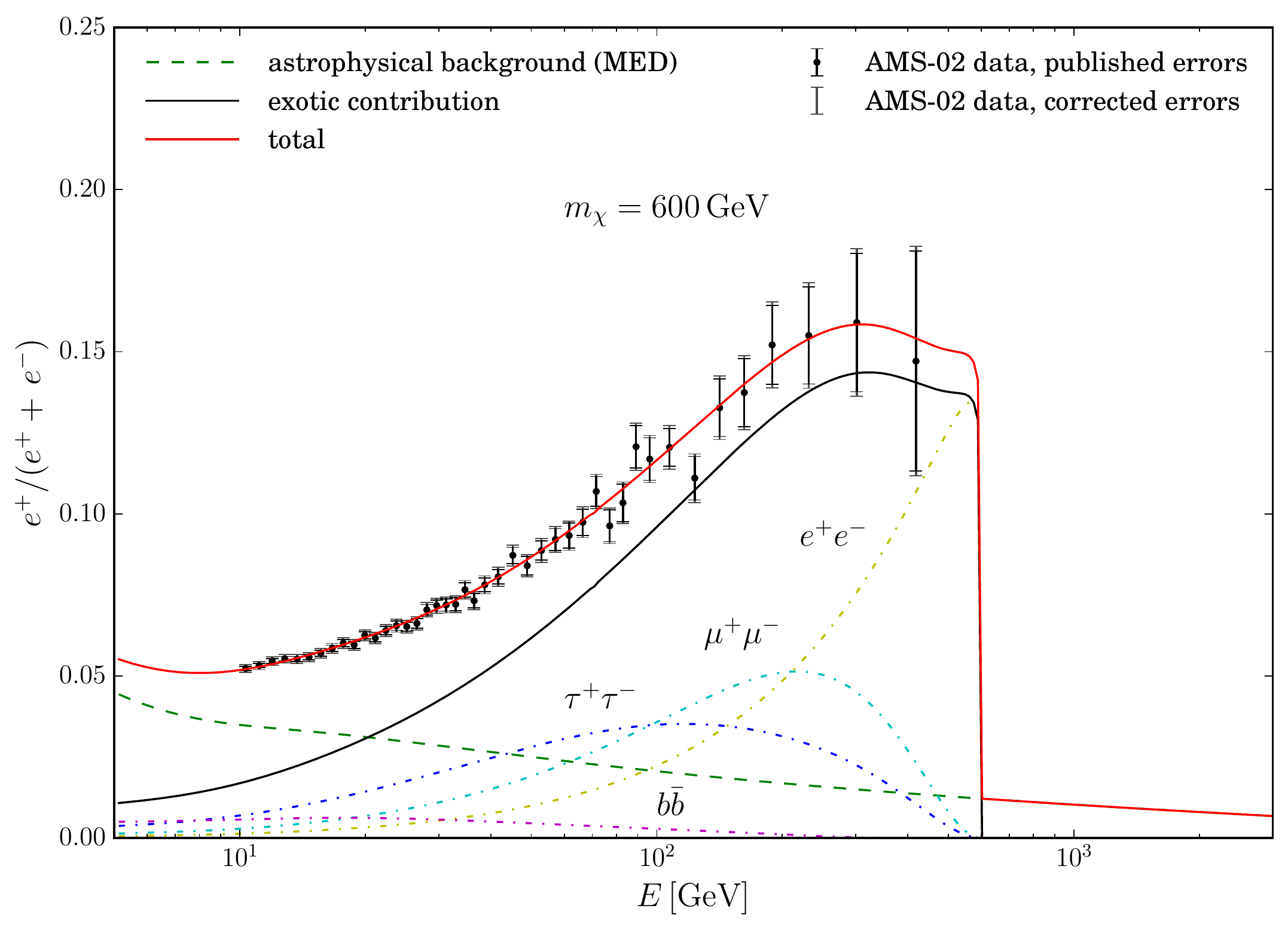}
\end{minipage}
\caption{
The WIMP annihilation signal and the astrophysical background contribute to the positron fraction, plotted here as a function
of energy and compared with AMS data~\cite{Accardo:2014lma}.
In the left panel, the case of a pure $b \bar{b}$ annihilation channel is featured. The best-fit values of $\langle \sigma v \rangle$
and DM mass $m_{\chi}$ are indicated. They correspond to a reduced chi-square of 0.65.
In the right panel, the possibility of a WIMP annihilating into lepton and $b \bar{b}$ pairs is featured to investigate how
mixed annihilation channels modify the best-fit DM parameters. The DM mass m$_{\chi}$ has been set equal to 600~GeV.
The cross section $\langle \sigma v \rangle$ and branching ratios are left free to vary until they best fit the positron fraction
data. The contribution of each channel to the signal is indicated. The branching ratio into $\tau^{+} \tau^{-}$ amounts to
50\% whereas the quark contribution is 20\%, with $\langle \sigma v \rangle = 1.11 \times 10^{-23}$ cm$^{3}$ s$^{-1}$.
These values yield an excellent fit with $\chi^{2}_{\rm dof} = 0.5$.
In both panels, the propagation parameters correspond to the {\sc Med} model.
Figure borrowed from~\cite{Boudaud:2014dta}.
{\hspace{12.0cm}}
}
\label{fig:1}
\vskip -0.5cm
\end{figure}
%

As a matter of fact, a positron excess was observed in 2008 above 10~GeV by the PAMELA satellite~\cite{Adriani:2008zr}.
More recently, the AMS-02 measurements~\cite{Accardo:2014lma} confirmed this anomaly up to 500~ GeV. As featured
in Fig.~\ref{fig:1}, the signal well exceeds the astrophysical background of secondary species. There must be a source of
additional positrons inside the Milky Way. An exciting possibility, which has been triggering a lot of activity and enthusiasm
in the particle physics community, is that this extra component is made of primary positrons produced by WIMP annihilation.
The fact that the anomaly is observed at high energy, where DM particles with a mass $\sim$ TeV are actually expected to
contribute, is a strong incentive in favor of the DM hypothesis.

High-energy positrons lose  rapidly their energy while they spiral inside the magnetic fields through which they propagate.
Those detected by AMS-02 must have been produced in the vicinity of the solar system in order to make it to the Earth. The
DM density in the solar neighborhood is $\sim$ 0.3 GeV cm$^{-3}$, so that the production term for primary positrons from
WIMP annihilation can be estimated. The only ingredient to be ajusted in the DM fit to the signal is the annihilation cross
section $\langle \sigma v \rangle$. In the Boudaud {\etal} analysis~\cite{Boudaud:2014dta}, the DM mass m$_{\chi}$ and
cross section $\langle \sigma v \rangle$ are fitted to the AMS-02 positron fraction. An agnostic scan over the various possible
annihilation channels is performed. For the first time, mixed channels are also considered as displayed in the right panel of
Fig.~\ref{fig:1}.
%
This analysis confirms that the annihilation cross section must be orders of magnitude larger than the Big Bang canonical value
of $3 \times 10^{-26}$ cm$^{3}$ s$^{-1}$. In the left panel of Fig.~\ref{fig:1} for instance, $\langle \sigma v \rangle$ needs
to be boosted by $3 \times 10^{4}$.
The leptonic channels are also disfavored by the data. Annihilation into tau pairs leads to a good fit, but muon pairs need to be
produced via a light mediator while electrons are always excluded. All the other channels have P-values in excess of 99\%.
An additional problem lies in the fact that for precisely those channels, DM particles are coupled to quarks as well as to gauge and
Higgs bosons, and eventually produce antiprotons through the hadronization of the final state species. Such couplings are severely
constrained under the penalty of overproducing antiprotons as shown by Cirelli {\etal}~\cite{Cirelli:2008pk} and confirmed by Donato
{\etal}~\cite{Donato:2008jk}. DM particles must be leptophilic on the one hand, but the AMS-02 data are not quite in favor of that
possibility on the other hand.
%
Another important piece of this puzzle comes from the observations of the gamma ray emission from dwarf spheroidal (dSph)
satellite galaxies of the Milky Way. These systems are known to be dominated by DM and would shine in the gamma ray sky should
DM strongly annihilate into high-energy photons. Recently, a comprehensive analysis by L{\'o}pez {\etal}~\cite{Lopez:2015uma}
indicates that the only viable channel that survives the Fermi/LAT constraint from dSph's and still produces a good fit to the AMS-02
positron fraction is DM annihilation via a mediator to 4 muons, or mainly to 4 muons in the case of multichannel combinations.

The excitement about the positron excess has by now receded. A more plausible explanation of the positron anomaly is based on pulsars.
These highly magnetized neutron stars do exist and some of them even lie in our vicinity. Their rotating magnetic field can accelerate
electron-positron pairs which are then released in the interstellar medium. The magnitude, spectral index and energy cut-off of the positron
spectrum produced by each pulsar can be separately adjusted to the AMS-02 data. The fits are so good that even a single object is enough
to explain the positron fraction, as demonstrated by~\cite{Linden:2013mqa,Cholis:2013psa,Boudaud:2014dta}.
Of course, nothing precludes DM to contribute to the positron excess. But this anomaly is no longer considered as a smoking gun
signature of the presence of WIMPs inside the Milky Way.

\newpage
\section{Antiprotons -- Has dark matter been discovered~?}
%
\begin{figure}[t!]
\begin{minipage}{6cm}
\centering
\includegraphics[width=6cm]{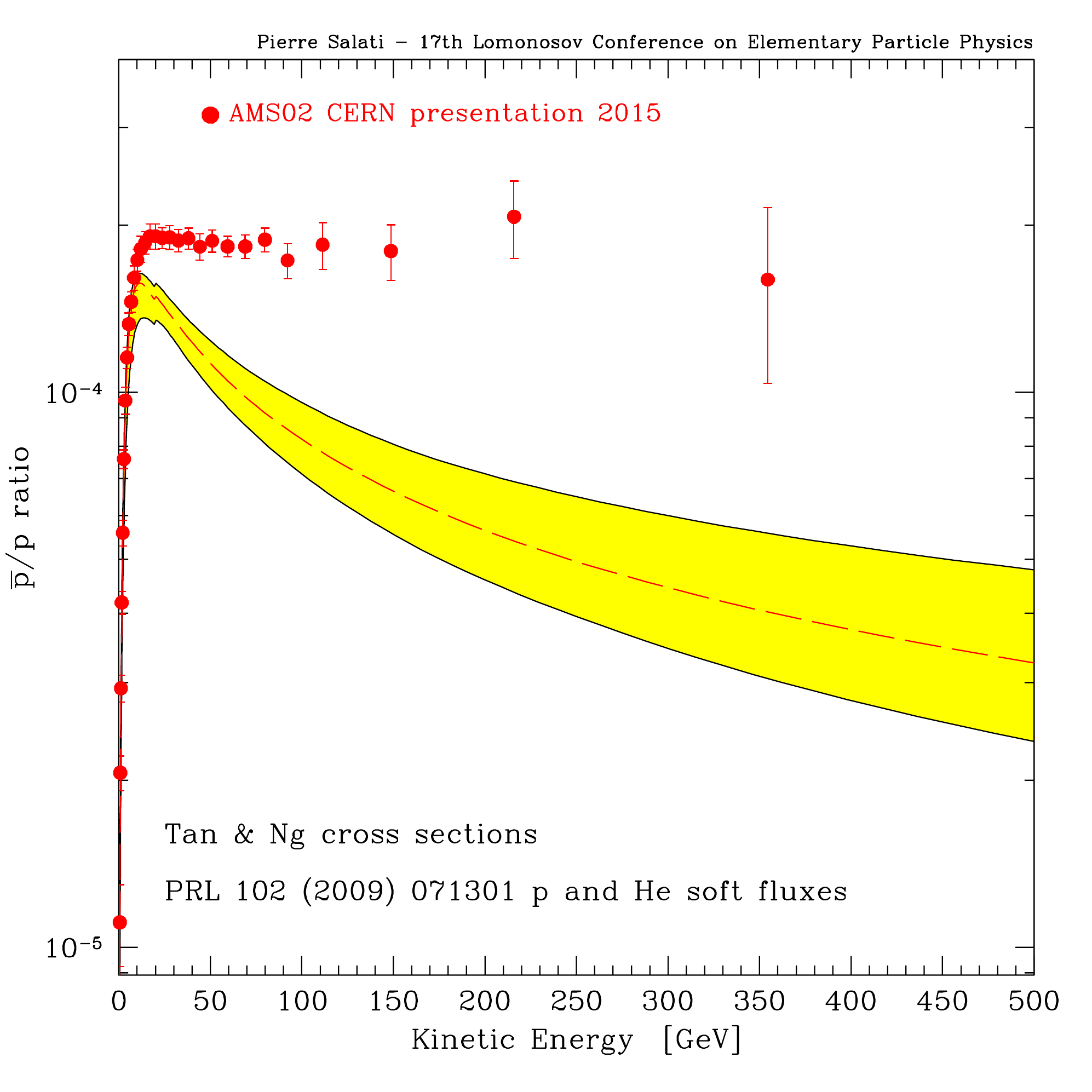}
\end{minipage}
\hfill
\begin{minipage}{6cm}
\centering
\includegraphics[width=6cm]{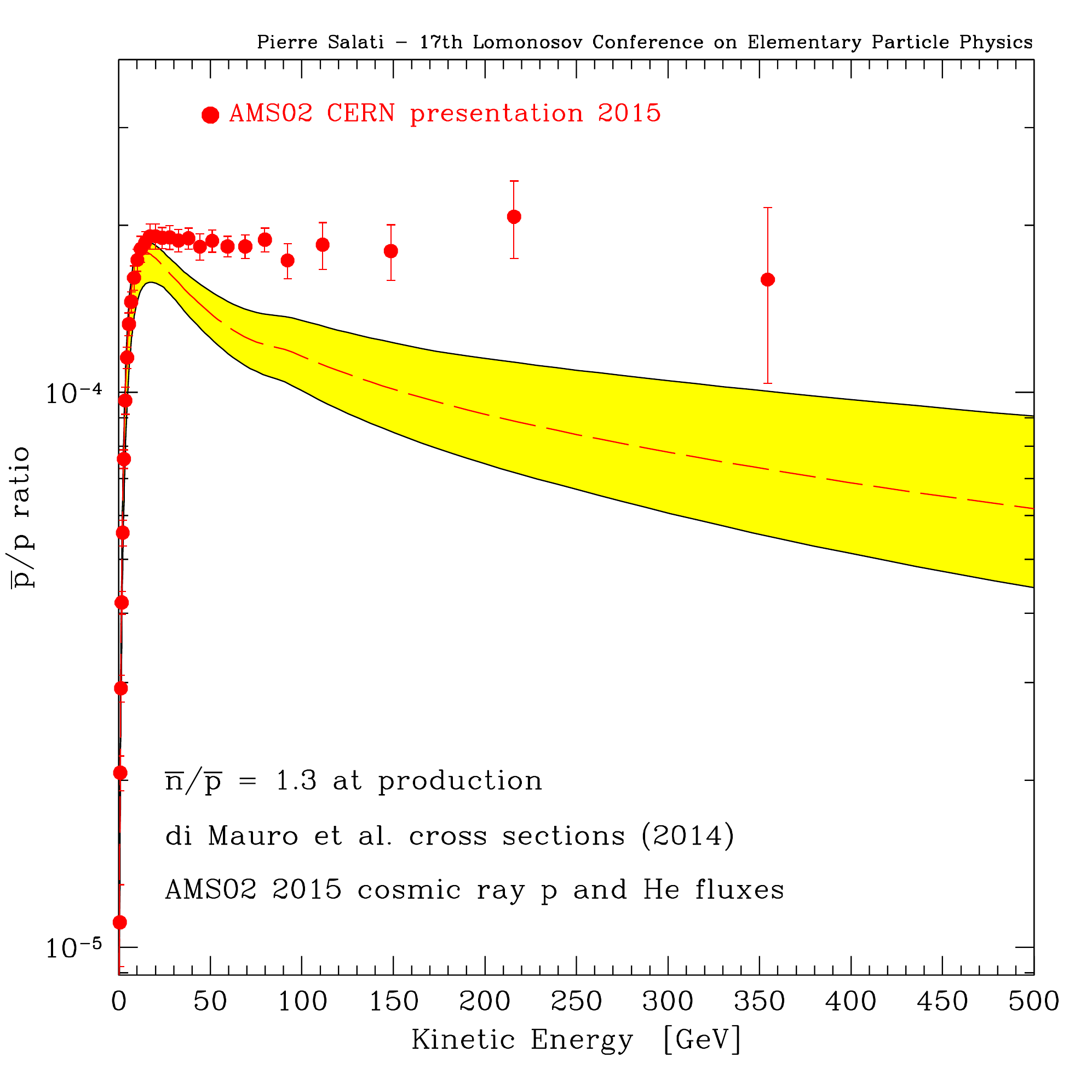}
\end{minipage}
\caption{
The antiproton-to-proton ratio ${\bar{\rm p}}/{\rm p}$ is plotted as a function of antiproton kinetic energy.
The theoretical prediction for the secondary antiproton background is featured by the red long-dashed curve,
whereas the preliminary results presented by the \AMS\ collaboration in April 2015 correspond to the red
dots~\cite{AMS2015}.
The yellow band encompasses the fluxes derived with cosmic ray propagation parameters compatible
with the B/C ratio~\cite{Maurin:2001sj}. It illustrates the uncertainty arising from cosmic ray transport in the Milky Way.
In the left panel, the background as calculated in 2008 by~\cite{Donato:2008jk} is presented. The gap between the
predictions and the data is obvious and may lead to the conclusion that an exotic antiproton component is necessary.
The tension between the expected background and the data is less severe in the right panel where new calculations
are featured~\cite{Giesen:2015ufa}. See text for details.
{\hspace{2.5cm}}
}
\label{fig:2}
\vskip -0.5cm
\end{figure}
%

A much more convincing signature for WIMPs would be an excess of antiprotons at high energy. No astrophysical objet
is known to produce antiprotons and to release them in the interstellar medium like pulsars do for positrons. In that context,
the preliminary measurement of the antiproton-to-proton ratio which the AMS-02 collaboration presented in April 2015~\cite{AMS2015}
aroused a renewed interest in WIMPs. As is clear in the left panel of Fig.~\ref{fig:2}, the gap between the astrophysical background
of secondary antiprotons and the data is significant. It could leave room for a primary component produced by DM annihilation.

But before hastily reaching the conclusion that a new anomaly has been brought to light, the astrophysical background needs to be
thoroughly investigated. As for positrons, it is produced by the collisions of high-energy protons and helium nuclei on interstellar
gas. The left panel of Fig.~\ref{fig:2} displays the result obtained in 2008 by Donato {\etal}~\cite{Donato:2008jk} whereas
the conspicuously larger background presented in the right panel is based on a recent analysis by Giesen {\etal}~\cite{Giesen:2015ufa}.
It leaves little room now for an exotic signal. As a matter of fact, three ingredients essential to the production of secondary antiprotons
have changed since 2008.
%
To commence, a secondary antiproton of energy $E$ is produced on average by a proton of energy $\eta E$ impinging on an interstellar
hydrogen atom at rest, with $\eta \sim 10$. The antiproton flux $\Phi_{\bar{\rm p}}(E)$ at energy $E$ is proportional to the proton
flux $\Phi_{\rm p}(\eta E)$ at energy $\eta E$. As the latter is a power law with spectral index $\alpha$, the ${\bar{\rm p}}/{\rm p}$ ratio
scales like $\eta^{- \alpha}$. The spectral index $\alpha$ has been decreasing for the past years and hardenings in the proton and helium
fluxes have recently been reported around 300~GeV~\cite{Aguilar:2015ooa,Aguilar:2015ctt}. Both observations lead to an increase of the
${\bar{\rm p}}/{\rm p}$ ratio.
%
Then, the antiproton production cross section in pp collisions ${d \sigma_{\rm p H \to \bar{p}}}/{dE_{\bar{\rm p}}}$ has been recently
reinvestigated. A new parameterization based on data from the BRAHMS and NA49 experiments is now available~\cite{diMauro:2014zea}.
It also induces a (modest) increase of the ${\bar{\rm p}}/{\rm p}$ ratio.
%
Last but not least, pp interactions could yield more antineutrons than antiprotons as suggested by~\cite{Kappl:2014hha}. The cross sections
of the reaction
${\rm p} {\rm p} \to \bar{\rm n} + {\rm X}$ and its flipped counterpart
${\rm n} {\rm p} \to \bar{\rm p} + {\rm X}$ are related by isospin symmetry. The NA49 experiment has measured a larger antiproton
multiplicity in np compared to pp collisions~\cite{Fischer:2003xh}. This observation can only be understood if pp interactions yield more
antineutrons than antiprotons, by a factor as large as 50\%.
%
\begin{figure}[t!]
\centering
\epsfig{file=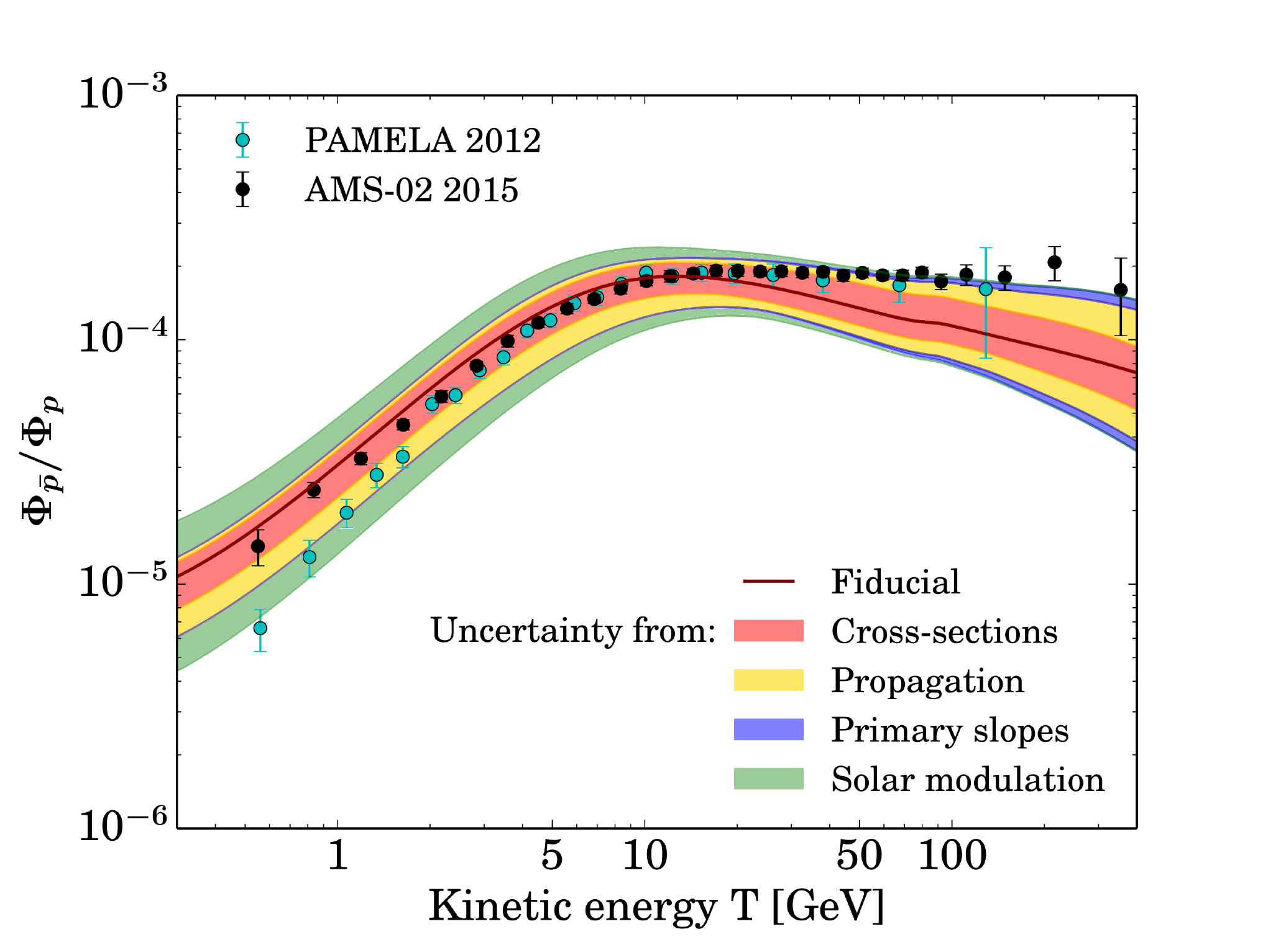 , width=10 cm}
\caption{
The combined total uncertainty on the predicted secondary ${\bar{\rm p}}/{\rm p}$ ratio, superimposed to the older
\PAMELA\ data~\cite{Adriani:2012paa} and the new \AMS\ data~\cite{AMS2015}. The curve labelled `fiducial'
assumes the reference values for the different contributions to the uncertainties: best fit proton and helium fluxes,
central values for the cross sections, {\sc Med} propagation and central value for the Fisk potential. We stress however
that the whole uncertainty band can be spanned within the errors.
Figure borrowed from~\cite{Giesen:2015ufa}.
{\hspace{4.5cm}}
}
\label{fig:3}
\vskip -0.5cm
\end{figure}
%

The calculations carried out by Giesen {\etal}~\cite{Giesen:2015ufa} are presented in Fig.~\ref{fig:3} where various sources of uncertainty
are plotted as colored bands. The data are still compatible with the astrophysical background, although they lie close to the upper
edge of the allowed region. Notice however that the fiducial prediction is based on the {\sc Med} cosmic ray propagation model which has
been derived from old B/C data. Forthcoming measurements of the B/C ratio may well induce a shift of the background region upward, as the
analysis by~\cite{Kappl:2015bqa} seems to indicate.

\section{Closing thoughts}

The cosmic ray positron anomaly has been confirmed by the AMS-02 collaboration. It is difficult to explain this excess solely by DM
anihilation. The regions in the WIMP mass and cross section parameter space have moved because measurements are now so much
precise. But very few of them survive after the Fermi/LAT constraints from dSph's are applied.
The most plausible explanation has to be found in nearby pulsars.
As regards antiprotons, the preliminary AMS-02 ${\bar{\rm p}}/{\rm p}$ ratio is compatible with a pure secondary component,
although the data are close to the upper edge of the expected background. To decide whether a DM signal is hidden, cosmic ray
propagation needs to be better constrained and the antiproton production cross sections in pp and NN collisions should be more
accurately measured.

\section*{Acknowledgments}

P.S. would like to thank the organizers of the 17th Lomonosov Conference on Elementary Particle Physics, in particular Alexander
Studenikin, for their warm hospitality and the friendly and inspiring atmosphere of the meeting. This work has been supported by
Institut universitaire de France.


\end{document}